\documentclass[aps,prl,reprint,superscriptaddress,nofootinbib]{revtex4-2}
\usepackage[utf8]{inputenc}
\usepackage{amsmath}
\usepackage{stix2}
\usepackage{bm}
\usepackage{xcolor}
\usepackage{graphicx}
\usepackage{booktabs}

\begin{document}

\title{Power-law growth of shift current with superlattice period in flat Chern bands}

\author{Nianlong \surname{Zou}}
\affiliation{Department of Physics and Astronomy, University of Tennessee, Knoxville, TN 37996, USA}

\author{Cheng Xu}
\affiliation{Max Planck Institute for Chemical Physics of Solids, 01187, Dresden, Germany}

\author{Ning Mao}
\affiliation{Max Planck Institute for Chemical Physics of Solids, 01187, Dresden, Germany}

\author{Yang Zhang}
%\email{yangzhang@utk.edu}
\affiliation{Department of Physics and Astronomy, University of Tennessee, Knoxville, TN 37996, USA}
\affiliation{Min H. Kao Department of Electrical Engineering and Computer Science, University of Tennessee, Knoxville, Tennessee 37996, USA}

\begin{abstract}
Quantum geometry governs a wide array of physical observables in topological quantum materials. In its ideal limit, quantum geometry yields exact bounds and analytical results for a growing class of observables. However, a comparable framework for the shift current remains elusive because it depends on geometric relations between multiple bands. In this work, we show that the shift current is proportional to the cyclotron shift, the displacement of the cyclotron center induced by Landau-level mixing. This mechanism yields a universal scaling law: the integrated weight and the peak magnitude scale as $l^{i+1-n}$ and $l^{2i+1-n}$ in the magnetic length or moir\'e period $l$, where $n$ is the momentum order of the symmetry-breaking perturbation and $\Delta E \propto l^{-i}$ is the level spacing. We verify the scaling law in exactly solvable chiral-$N$ Landau levels under a uniform magnetic field and in Schr\"odinger and Dirac moir\'e skyrmion crystals, which capture the flat Chern bands of twisted MoTe$_2$ and twisted bilayer graphene. For a Schr\"odinger flat band with a distorted skyrmion texture, the integrated weight grows linearly with the moir\'e period and the resonant peak grows cubically.
\end{abstract}

\maketitle

Flat Chern bands in moir\'e materials have emerged as a versatile platform for studying quantum geometric effects. In these systems, the suppression of kinetic energy allows quantum geometry to govern a wide variety of physical properties, ranging from superfluid weight~\cite{Peotta2015, Hu2019, Xie2020} and optical responses~\cite{Ahn2020,Chaudhary2022,Avdoshkin2025,Mitscherling2025} to static structure factors and dielectric properties~\cite{Yu2025a,Shavit2025,Onishi2024a,Onishi2024,Onishi2025,Passos2026}. Much of this progress relies on \emph{ideal quantum geometry}, the condition under which the inequalities bounding geometric quantities by topological quantities (such as the quantum metric by the Berry curvature) are saturated~\cite{Roy2014}. In this ideal limit, otherwise complicated geometric quantities become analytically tractable, giving rise to universal bounds and exact expressions for a growing class of physical observables~\cite{Onishi2024,Passos2026}.

The shift current describes the generation of a dc photocurrent in noncentrosymmetric materials and has long been pursued as a photovoltaic mechanism beyond the $p$--$n$ junction paradigm~\cite{vonBaltz1981,Sipe2000,Young2012,Morimoto2018}. The shift current is likewise strongly influenced by band geometry, with particular attention devoted to its enhancement in topological gapless states and flat bands~\cite{Morimoto2018,Ahn2020,Avdoshkin2025,Alexandradinata2024,Chaudhary2022,Zhang2018,Atlam2025,Zou2025,wu2024}. However, unlike the above observables, the shift current is an intrinsically multiband response governed by the multistate theory of quantum geometry rather than by single-band geometric quantities~\cite{Avdoshkin2025,Mitscherling2025,Alexandradinata2024}. The natural question is whether the shift current admits a similarly exact description rooted directly in ideal quantum geometry.

Landau levels (LLs) provide the natural arena to realize ideal quantum geometry exactly~\cite{GMP1986,Haldane2011,Wang2021,Ledwith2023,Liu2025,Li2026}. Their nonlinear optics has recently moved into focus, from third-order nonlinearities~\cite{Yao2012,Avetissian2016,Yumoto2018} and second-harmonic generation~\cite{Lu2023} to the shift current~\cite{Morimoto2018,Mao2026,Bednik2024}. However, in existing works, the macroscopic responses are often evaluated case-by-case, and a unified picture of the shift current in the LL basis remains to be established. In this work, based on the decomposition of the position operator into guiding-center and cyclotron coordinates, we show that the shift current in LLs and ideal flat Chern bands is proportional to the displacement of the cyclotron center induced by LL mixing. Because ideal quantum geometry dictates that the magnetic length $l_B$ (or moir\'e period $l_m$) is the system's sole characteristic length scale, this geometric response follows a universal power-law scaling with respect to $l_B$ ($l_m$). The exponent is fixed by the momentum order $n$ of the symmetry-breaking perturbation and by the scaling of the level spacing. We then numerically verify these analytical predictions in exactly solvable chiral-$N$ LL models, where the integrated shift current crosses over from an $l_B^{0}$ plateau to an $l_B^{-2N}$ decay. In moir\'e skyrmion crystals of Schr\"odinger and Dirac electrons, the peak shift current of a given LL transition scales as $l_m^{5-n}$ and $l_m^{3-n}$, respectively.

\textbf{Shift current in the Landau-level basis.---} The shift conductivity of a two-dimensional system is
\begin{equation}\label{sc}
  \sigma^{\alpha \beta \gamma} = \frac{\pi e^3}{\hbar^2} \int \frac{\mathrm{d} \bm{k}}{(2 \pi)^2} \sum_{n, m} f_{nm} I_{mn}^{\alpha\beta\gamma} \delta (\omega_{nm} - \omega),
\end{equation}
where $\alpha, \beta, \gamma$ are Cartesian indices, $\omega$ denotes the frequency of the incident light, and \(f_{nm} = f_n - f_m\) with \(f_n\) and \(f_m\) being the Fermi distribution functions. The integrand $I_{mn}^{\alpha\beta\gamma}=\text{Re}[ (\mathcal{R}^{\alpha,\beta}_{nm}-\mathcal{R}^{\alpha,\gamma}_{mn})r_{mn}^{\gamma}r_{nm}^{\beta}]$, with the shift vector $\mathcal{R}_{nm}^{\alpha,\beta}$ defined as $\mathcal{R}_{nm}^{\alpha, \beta} (\bm{k}) = \partial_{k_\alpha} \arg r_{nm}^\beta (\bm{k}) - (\mathcal{A}_n^\alpha (\bm{k}) -\mathcal{A}_m^\alpha (\bm{k}))$ where \(\mathcal{A}^{\alpha}_{n}\) and $r_{nm}^\beta$ are the intraband and interband Berry connections, respectively. Physically, this equation illustrates that the macroscopic photocurrent arises from a spatial displacement of the electron during the optical transition, captured by the shift vector and weighted by the transition probability $r_{mn}^{\gamma}r_{nm}^{\beta}$.

\begin{figure}
    \includegraphics[width=\linewidth]{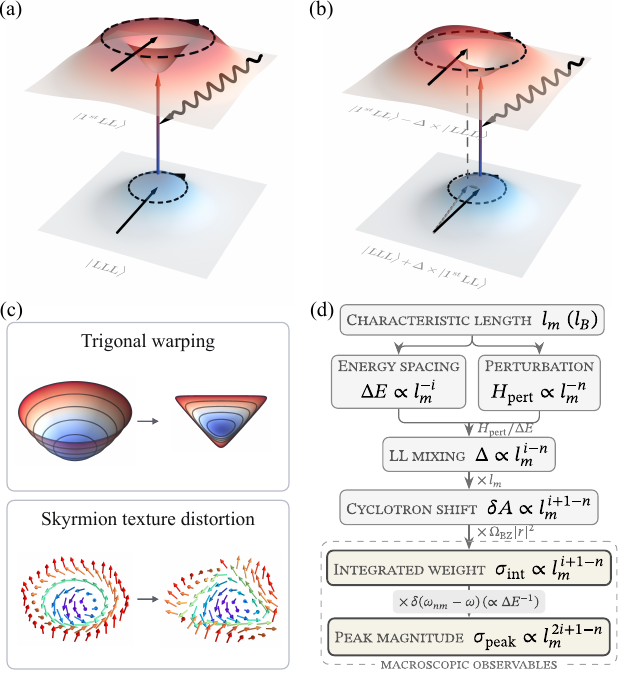}
    \caption{(a), (b) Schematics of the shift-current response for (a) pure LLs and (b) LLs with perturbative mixing. Dashed circles and black arrows denote the cyclotron orbits and their centers, respectively. (c) Two symmetry-breaking channels: trigonal warping of the dispersion and distortion of the skyrmion texture. (d) Flow diagram from the characteristic length $l_m$ ($l_B$) through the LL mixing $\Delta$ and the cyclotron shift $\delta A$ to the integrated weight and the peak magnitude of the shift current. The level-spacing exponent is $i=2$ ($i=1$) for the Schr\"odinger (Dirac) SkX model, and $n$ is the momentum order of the perturbation.}
    \label{fig1}
\end{figure}

The evaluation of the shift current in the LL basis is greatly simplified by the decomposition of the position operator, $\hat{\bm r}=\hat{\bm R}+\hat{\bm\eta}$. The guiding-center coordinate $\hat{\bm R}$ acts within each degenerate LL manifold, while the cyclotron coordinate $\hat{\bm\eta}$ mediates inter-LL transitions. Accordingly, the interband and intraband Berry connections originate respectively from the cyclotron and guiding-center motions. Expressing $\hat{\bm{\eta}}$ in terms of ladder operators reveals that the interband connection between LLs, $\bm{r}_{nm}^{\mathrm{LL}}$, is $k$-independent, connects only adjacent LLs, and scales proportionally with $l_B$\cite{SM}. In contrast, the intraband connection is determined exclusively by the guiding center $\hat{\bm{R}}$. Because the guiding-center algebra is LL-independent, every level shares the identical connection $\mathcal{A}^{\mathrm{LL};\alpha}(\bm{k}) = \frac{l_B^2}{2}\epsilon_{\alpha\beta}k_\beta$.  As illustrated in Fig.~\ref{fig1}(a), an optical transition excites only the cyclotron motion while leaving the guiding center unchanged. Consequently, the shift vector, which in this basis reduces to the difference in guiding-center positions, vanishes and the shift current is zero, consistent with the rotational symmetry of the unperturbed LL problem. A finite shift current emerges only when LL mixing distorts the ideal cyclotron orbits and induces a displacement of the cyclotron center [Fig.~\ref{fig1}(b)]. Explicitly, for mixed LL states with $k$-independent mixing coefficients~\cite{SM}, the shift vector can be expressed as
\begin{equation}\label{shiftvec}
\mathcal{R}_{nm}^{\alpha,\beta} = \delta A_m^\alpha - \delta A_n^\alpha,
\end{equation}
where we refer to $\delta A_n^\alpha$ as the \emph{cyclotron shift}, defined by $\delta A_n^\alpha \equiv \langle \Psi_n|\hat{\eta}^\alpha|\Psi_n\rangle$, with $|\Psi_n\rangle$ denoting a mixed LL state.

We focus on the regime in which the wave functions remain close to the ideal LL states, such that the LL mixing can be treated perturbatively, $|\Psi_n \rangle = | n \rangle + \Delta \sum_{i \neq n} c_{n i} | i \rangle$, with $| n \rangle$ the (relativistic) LL.
In this regime, the cyclotron shift is linear in the mixing strength $\Delta$. At fixed $\Delta$, increasing the magnetic length $l_B$ proportionally enlarges the cyclotron orbit and hence the cyclotron shift, giving $\delta A_n^\alpha \propto \Delta l_B$. Since the cyclotron-shift difference $\delta A_m^\alpha-\delta A_n^\alpha$ is therefore $\mathcal{O}(\Delta)$, the leading-order shift current is obtained by evaluating all other factors at zeroth order. In particular, the interband dipoles can be replaced by their bare LL values, $r_{mn}^{\mathrm{LL}}$. The frequency-integrated shift current between LLs $n$ and $m$, $W_{mn}^{\alpha\beta\gamma}\equiv\int d\omega\,\sigma_{mn}^{\alpha\beta\gamma}$, is thus given by
\begin{equation}
W_{mn}^{\alpha \beta \gamma} \approx \frac{2\pi e^3g_\text{LL}}{\hbar^2} (\delta A_m^\alpha - \delta A_n^\alpha) \operatorname{Re} [ r_{mn}^{\mathrm{LL};\gamma} r_{nm}^{\mathrm{LL};\beta} ],\label{scLL}
\end{equation}
where $g_\text{LL}=1/(2\pi l_B^2)$ is the LL degeneracy per unit area. This linear dependence on the characteristic wave-function size is consistent with previous bounds on shift-current generation in extended systems~\cite{Tan2019}.

%Because the cyclotron shift is bounded by the cyclotron radius and scales as $\delta A_n^\alpha\sim\Delta l_B$, Eq.~\eqref{scLL} shows that the shift current reaches only an $\mathcal{O}(\Delta)$ fraction of the geometric upper bound set by the spatial extent of the wave functions~\cite{Tan2019}.

\textbf{Universal scaling law.---} Since the shift current in the flat-band limit is governed by the quantum geometry rather than the microscopic details of the Hamiltonian, the perturbative LL-mixing mechanism naturally extends to flat Chern bands in periodic moir\'e superlattices. Under ideal quantum geometry, the Berry curvature is directly inherited from the LL Berry connection, $\Omega_{\bm{k}} = \nabla_{\bm{k}} \times \bm{\mathcal{A}}^{\mathrm{LL}} = l_B^2$. For such a uniform $\Omega_{\bm{k}}$, the quantized Chern number $C$ dictates that $\Omega_{\bm{k}} = 2\pi C/\Omega_{\mathrm{BZ}} \propto l_m^2$, so that the emergent magnetic length $l_B$ is proportional to the moir\'e period $l_m$. The LL degeneracy $g_\text{LL}$ is likewise replaced by the Brillouin-zone integration measure $\Omega_{\mathrm{BZ}}/(2\pi)^2$.

To demonstrate the scaling law explicitly, we consider a generic moir\'e skyrmion (SkX) Hamiltonian, $H=T+J_0\hat{\bm{s}}\cdot\hat{\bm{n}}(\bm{r})$, where $\hat{\bm{s}}$ is the spin operator and $\hat{\bm{n}}(\bm{r})$ denotes the moir\'e skyrmion texture. We examine two paradigmatic kinetic terms $T$~\cite{Paul2023,Guan2023}: the Schr\"odinger SkX model with $T=\bm{p}^2/(2m)$ and the Dirac SkX model with $T=v_F\boldsymbol{\sigma}\cdot\bm{p}$, where $\boldsymbol{\sigma}$ acts on a pseudospin. These models capture the low-energy physics of representative moir\'e systems, including twisted $\mathrm{MoTe}_2$ for the Schr\"odinger case~\cite{Wu2019a,Morales-Duran2024a,Shi2024} and twisted bilayer graphene for the Dirac case~\cite{Guerci2025a,Khalaf2021}.

In the adiabatic limit ($J_0 \rightarrow \infty$), the electron spin aligns locally with the magnetic texture. A local SU(2) rotation $\mathcal{U}(\bm{r})$, defined by $\mathcal{U}^{\dagger}[\hat{\bm{s}}\cdot\hat{\bm{n}}(\bm{r})]\mathcal{U}=\hat{s}_z$, aligns the spin quantization axis with $\hat{\bm{n}}(\bm{r})$ and projects the system onto the low-energy adiabatic manifold, generating an emergent vector potential $\bm{A}$~\cite{Paul2023}. Thus, the skyrmion texture acts as an effective magnetic field and realizes the ideal quantum geometry of LLs. Introducing dimensionless coordinates $\tilde{\bm{r}}=\bm{r}/l_m$ and $\tilde{\bm{p}}=\bm{p}l_m/\hbar$, the resulting adiabatic Hamiltonians take the form:
\begin{align}
    H_{\mathrm{ad}}^{\mathrm{Schr}}  & = \frac{\hbar^2}{m l_m^2} \left[ \frac{1}{2}(\tilde{\bm{p}} - \tilde{\bm{A}})^2 + \tilde{V} \right], \label{eq:Had_free} \\
    H_{\mathrm{ad}}^{\mathrm{Dirac}} & = \frac{\hbar v_F}{l_m} \left[ \boldsymbol{\sigma} \cdot (\tilde{\bm{p}} - \tilde{\bm{A}}) \right], \label{eq:Had_dirac}
\end{align}
where $\tilde{\bm{A}}=i\mathcal{U}^{\dagger}\tilde{\nabla}\mathcal{U}$ and $\tilde{V} \propto (\tilde{\nabla}\hat{\bm{n}})^2$ are the dimensionless emergent vector and scalar potentials. The overall prefactors set the level spacing $\Delta E \propto l_m^{-i}$, with $i=2$ and $i=1$ for the Schr\"odinger and Dirac SkX models, respectively.

However, the ideal unperturbed SkX models carry no shift current due to a generalized inversion symmetry $\tilde{\mathcal{P}} = \mathcal{P}\otimes\mathcal{U}_z$, where $\mathcal{P}$ is spatial inversion and $\mathcal{U}_z$ is a $\pi$ spin rotation about the $z$-axis. Since $\mathcal U_z$ acts only in spin space, $\tilde{\mathcal P}$ constrains the response function in the same manner as ordinary inversion, forcing the shift current to vanish. To activate a finite shift current, we consider two distinct $\tilde{\mathcal{P}}$-breaking perturbations [Fig.~\ref{fig1}(c)]. The first introduces a warping term $H_{\mathrm{warp}} \propto k^n$ to the kinetic energy. Under minimal coupling, $\bm{k}\rightarrow(\tilde{\bm p}-\tilde{\bm A})/l_m$, so that $H_{\mathrm{warp}}^{(n)}\propto l_m^{-n}$. The second directly distorts the skyrmion texture $\hat{\bm n}(\bm r)$ through a phase shift $\varphi$ of its helical components~\cite{SM}, modifying only the dimensionless potentials $\tilde{\bm A}$ and $\tilde V$ in Eqs.~\eqref{eq:Had_free} and \eqref{eq:Had_dirac}. Its matrix elements therefore inherit the overall energy scaling of the adiabatic Hamiltonian, i.e., $n=2$ for the Schr\"odinger model and $n=1$ for the Dirac model. In both cases, the resulting LL-mixing coefficient is set by the ratio of the perturbation matrix element to the LL spacing, $\Delta\sim H_{\mathrm{pert}}/\Delta E$ $\propto l_m^{i-n}$. Combining $\Delta$ with the cyclotron shift $\delta A \propto \Delta l_m$, the Brillouin-zone measure $\Omega_{\mathrm{BZ}} \propto l_m^{-2}$, and the bare dipole $|r^{\mathrm{LL}}|^2 \propto l_m^{2}$ yields the integrated weight $W \propto l_m^{i+1-n}$. The peak magnitude carries an additional factor from the delta function, $\delta(\omega_{nm}-\omega) \propto \Delta E^{-1} \propto l_m^{i}$, and scales as $\sigma_{\mathrm{peak}} \propto l_m^{2i+1-n}$, as summarized in the flow diagram of Fig.~\ref{fig1}(d). This scaling of the peak magnitude assumes a broadening that scales with the level spacing. If the broadening $\Gamma$ is instead fixed, then $\sigma_{\mathrm{peak}} \approx W/\Gamma$, and the peak magnitude follows the integrated-weight scaling $\propto l_m^{i+1-n}$ once $\Gamma$ exceeds $\Delta E$.

\begin{figure}
    \includegraphics[width=\linewidth]{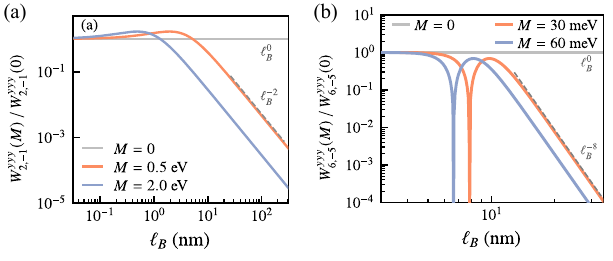}
    \caption{
    (a) Integrated shift current $|W^{yyy}_{2,-1}(M)/W^{yyy}_{2,-1}(0)|$ for massive and massless Dirac models (chiral-$N$ model with $N=1$) under a uniform magnetic field, for $M=0$, 0.5, and 2~eV. (b) Integrated shift current $|W^{yyy}_{6,-5}(M)/W^{yyy}_{6,-5}(0)|$ for tetralayer rhombohedral graphene (chiral-$N$ model with $N=4$) under a uniform magnetic field, for $M=0$, 30, and 60~meV. Dashed lines mark the $l_B^{0}$ and $l_B^{-2N}$ asymptotes, and the dips in (b) are sign changes of the cyclotron-shift difference.}
    \label{fig2}
\end{figure}

\textbf{Chiral-$N$ model under a uniform magnetic field.---} We first evaluate an exactly solvable two-band chiral-$N$ model~\cite{Min2008},
\begin{equation}
    \left( \begin{array}{cc}
     M & \gamma (\pi_x - i \pi_y)^N\\
     \gamma (\pi_x + i \pi_y)^N & - M
   \end{array} \right),
\end{equation}
where $M$ is the mass term, with the characteristic energy scale given by $\omega_N \equiv \gamma(\sqrt{2}\hbar/l_B)^N$. This framework yields relativistic LLs characterized by two-component spinors, capturing both the gapped Dirac equation ($N=1$) and tetralayer rhombohedral graphene (RHG, $N=4$) under a uniform magnetic field~\cite{Slizovskiy2019,Koshino2011,Jung2013}. A warping term of momentum order $n \le N$ does not generate a level-dependent cyclotron shift and produces no shift current, so the minimal activating perturbation has order $n=N+1$, i.e., $n=2$ for the Dirac model and $n=5$ for RHG~\cite{SM}. For this perturbation, the cyclotron shift scales as $\delta A \propto \Delta l_B \propto l_B^{N+1-n} = l_B^{0}$,
so that the shift vector is independent of $l_B$.

% translates all guiding centers by the same amount and produces no shift current, so the minimal activating perturbation has order $n=N+1$, i.e., $n=2$ for the Dirac model and $n=5$ for RHG~\cite{SM}. For this perturbation, the cyclotron shift scales as $\delta A \propto \Delta l_B \propto l_B^{N+1-n} = l_B^{0}$, so that the shift vector is independent of $l_B$, and for $N=1$ it is also independent of $M$~\cite{SM}.}

Consequently, the frequency-integrated shift current is dictated entirely by the interband Berry connection, leading to a crossover between two scaling regimes [Fig.~\ref{fig2}]. In the chiral regime ($\omega_N \gg M$), the system exhibits an approximate chiral symmetry, causing the spinor to be equally weighted on both sublattices. The interband Berry connection between spinors is then proportional to that of simple LLs, yielding an optical transition dipole that scales directly with the cyclotron radius ($\vert r \vert \propto l_B$). This compensates the LL degeneracy $g_\text{LL}\propto l_B^{-2}$ and yields a universal plateau $W \propto l_B^0$ for both models. Conversely, in the large-gap regime ($\omega_N \ll M$), the mass gap polarizes the occupied and empty states onto opposite sublattices, which suppresses the dipole matrix element to $\vert r \vert \propto l_B^{1-N}$. Therefore, the shift current decays as $W \propto \vert r \vert^2 l_B^{-2} \propto l_B^{-2N}$. This unified scaling yields an $l_B^{-2}$ decay for the Dirac fermion [$N=1$, Fig.~\ref{fig2}(a)] and a much steeper $l_B^{-8}$ drop for tetralayer RHG [$N=4$, Fig.~\ref{fig2}(b)].

\begin{figure}
    \includegraphics[width=\linewidth]{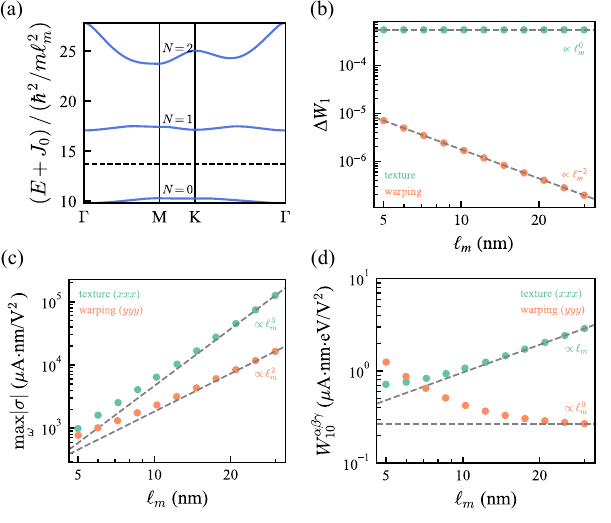}
    \caption{
    (a) Band structure of the Schr\"odinger SkX model in the adiabatic limit, showing the pseudo-LL ladder $N=0$, 1, 2. (b) LL-weight deficit $\Delta W_1 = 1-W_1$ of the $N=1$ miniband versus $l_m$ under the texture and warping perturbations. (c) Peak magnitude $\max_\omega|\sigma|$ and (d) frequency-integrated weight $W_{10}$ of the shift current for the $0\to1$ transition versus the moir\'e period $l_m$ under the two perturbations. Dashed lines are the predicted power laws.
    }
    \label{fig3}
\end{figure}

\textbf{Schr\"odinger skyrmion model.---} We now verify this framework numerically for the Schr\"odinger SkX model. In the deep adiabatic regime, the lowest minibands form an equally spaced ladder of pseudo-LLs [Fig.~\ref{fig3}(a)], with spacing set by the emergent cyclotron energy $\hbar\omega_c \propto l_m^{-2}$. To activate the shift current from the $\tilde{\mathcal{P}}$-invariant unperturbed state, we introduce the two aforementioned symmetry-breaking channels: the texture distortion by the phase $\varphi$ (acting as an $n=2$ perturbation) and a trigonal warping $H_3 \propto p_x^3 - 3 p_x p_y^2$ (an $n=3$ perturbation).

The mixing parameter $\Delta$ can be extracted directly from the wave functions. Within skyrmion lattice model~\cite{Paul2023,Reddy2024,xu2025multiple}, we calculate the LL weight $W_N$, whose deficit $\Delta W_N = 1 - W_N$ measures the mixing magnitude $\Delta^2$. As shown in Fig.~\ref{fig3}(b), the deficit scales as $\Delta W_1 \propto l_m^{0}$ under the $\varphi$ perturbation and $\Delta W_1 \propto l_m^{-2}$ under trigonal warping, in agreement with the predicted $\Delta \propto l_m^{2-n}$ scaling. The flow diagram of Fig.~\ref{fig1}(d) then predicts a peak magnitude $\sigma_{\mathrm{peak}} \propto l_m^{5-n}$ and an integrated weight $W \propto l_m^{3-n}$. Remarkably, the peak shift current grows with the moir\'e period, as $l_m^{3}$ for the $\varphi$ perturbation and as $l_m^{2}$ for warping, even though the perturbation matrix elements themselves decrease as $l_m^{-2}$ and $l_m^{-3}$. Our numerical results [Fig.~\ref{fig3}(c)] converge onto these power laws at large $l_m$, where the adiabatic condition $J_0/\hbar\omega_c \gg 1$ holds, and the integrated weights approach the companion laws $l_m^{1}$ and $l_m^{0}$ [Fig.~\ref{fig3}(d)]. Since $\hbar\omega_c$ falls to $\approx1.5$~meV at $l_m=25$~nm for the parameters used here~\cite{SM}, below typical broadenings, the integrated weight $W$ rather than the peak is the experimentally robust quantity in this regime. Expressed in terms of the twist angle $\theta \propto 1/l_m$ of twisted $\mathrm{MoTe}_2$, the $\varphi$ channel yields $\sigma_{\mathrm{peak}} \propto \theta^{-3}$.

\begin{figure}
    \includegraphics[width=\linewidth]{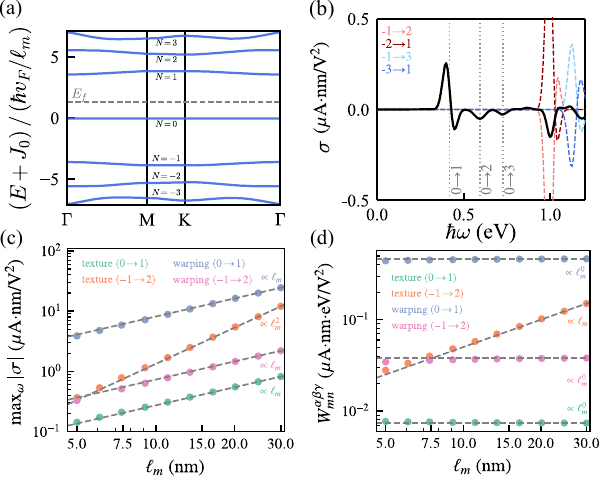}
    \caption{
        (a) Band structure of a Dirac electron coupled to a skyrmion texture in the adiabatic limit, showing the relativistic ladder $N=0,\pm1,\pm2,\pm3$.
        (b) Shift conductivity spectrum (solid) and band-resolved contributions of the cross transitions $-1\to2$, $-2\to1$, $-1\to3$, and $-3\to1$ (dashed), showing the chiral-symmetry-induced cancellation of cross transitions (class i), the survival of zero-mode transitions (class ii), and the self-conjugate diagonal transitions (class iii).
        (c) Peak magnitude and (d) frequency-integrated weight of the band-resolved shift current for the $0\to1$ and $-1\to2$ transitions versus the moir\'e period $l_m$ under the texture and warping perturbations. Dashed lines are the predicted power laws.
    }
    \label{fig4}
\end{figure}

\textbf{Dirac skyrmion model.---} Next, we turn to the numerical results of the Dirac SkX model. In the adiabatic regime, the minibands reproduce a relativistic LL ladder, $E_N \propto \mathrm{sgn}(N)\sqrt{|N|}\,\hbar\omega_0$ with $\hbar\omega_0 \propto l_m^{-1}$ [Fig.~\ref{fig4}(a)]. Again, the shift conductivity is driven by two types of perturbations: the same texture distortion $\varphi$, which acts as an $n=1$ perturbation in this case, and the quadratic warping, $H_{\mathrm{warp}}\propto[(p_x^2-p_y^2)\tau_x+2p_xp_y\tau_y]$, which acts as an $n=2$ perturbation. Crucially, the adiabatically projected Hamiltonian, including both perturbations, is purely off-diagonal in the sublattice index. This enforces an emergent \emph{chiral symmetry} that pairs the $\pm N$ levels, which is violated only by nonadiabatic corrections of order $\hbar\omega_0/J_0$.

This symmetry organizes the optical transitions into three distinct classes [Fig.~\ref{fig4}(b)]. (i) A cross transition $-n\to m$ is spectrally degenerate with its chiral conjugate $-m\to n$. The chiral map forces $\sigma_{-n\to m} = -\sigma_{-m\to n}$, causing the two large, adiabatic-regular members to cancel pairwise and leaving only a residual peak of order $\hbar\omega_0/J_0$. (ii) The zero mode is chiral-invariant, and $0\to n$ has no conjugate partner, so this family escapes the cancellation and dominates the spectrum. (iii) The diagonal transitions $-n\to n$ are self-conjugate under the chiral map, so the same antisymmetry forces $\sigma_{-n\to n} = -\sigma_{-n\to n} = 0$ in the chiral limit.

% The scaling laws of the band-resolved contributions [Figs.~\ref{fig4}(c,d)] again follow Fig.~\ref{fig1}(d) with $i=1$, but reveal an asymmetry between the two perturbations. For the SkX texture perturbation, which acts as an $n=1$ perturbation, the individual cross transitions (such as $-1\to 2$) obey the generic laws $\sigma_{\mathrm{peak}} \propto l_m^{3-n} = l_m^{2}$ and $W$ $\propto l_m$, independent of the exchange coupling $J_0$\red{. However, their net contribution to the spectrum is reduced by the pairwise cancellation of class (i)}. For the quadratic warping, which acts as an $n=2$ perturbation, the zero mode remains protected \red{because the warping annihilates the $N=0$ state~\cite{SM}. The} upward mixing of higher levels \red{still} renders \red{the} $0\to 1$ transition adiabatic-regular, yielding the generic $n=2$ limits $\sigma_{\mathrm{peak}} \propto l_m^1$ and a constant integrated weight. \red{By contrast, the warping contribution of the $-1\to 2$ transition scales as $\sigma_{\mathrm{peak}} \propto l_m^{0}$ and $W \propto l_m^{-1}$, one power of $l_m$ below the generic $n=2$ law.}

The scaling laws of the cross transitions [such as $-1\to 2$ in Fig.~\ref{fig4}(c,d)] again follow Fig.~\ref{fig1}(d) with $i=1$. For the $n=1$ texture perturbation, the band-resolved signals strictly obey $\sigma_{\mathrm{peak}} \propto l_m^{2}$ and $W \propto l_m^1$. For the $n=2$ warping perturbation, the scaling is reduced by one power of $l_m$, yielding $\sigma_{\mathrm{peak}} \propto l_m$ and $W \propto l_m^0$. However, their net contribution to the total spectrum is heavily suppressed by pairwise cancellation: the surviving fraction of the band-resolved signal is set by the nonadiabatic scale $\hbar\omega_0/J_0$, which measures the strength of chiral-symmetry-breaking corrections.

The situation for the $0\to 1$ transition is entirely different and depends on the nature of the perturbation. For the quadratic warping, this transition is adiabatic-regular; the warping perturbation mixes the $N=0$ state with higher levels~\cite{SM}, fulfilling the generic $n=2$ scaling law with $\sigma_{\mathrm{peak}} \propto l_m^1$ and a constant integrated weight. By contrast, for the $\varphi$ perturbation, the strict-adiabatic response vanishes due to the topological protection of the zeroth LL. Its surviving signal is thus entirely generated by finite-$J_0$ nonadiabatic effects. As derived via a Schrieffer--Wolff transformation~\cite{SM}, the off-diagonal gauge-field components generate an effective Dirac mass $M_{\mathrm{eff}}(\bm{r}) = -\pi (v_F^2/J_0)\,\rho_{\mathrm{top}}(\bm{r})$. This mass term breaks the chiral symmetry, lifting the protection of the zero mode and yielding a distinct nonadiabatic scaling: the integrated weight is independent of $l_m$ ($W \propto J_0^{-1}$), while the peak conductivity scales as $\sigma_{\mathrm{peak}} \propto l_m/J_0$.

\textbf{Conclusion.---} In conclusion, we have shown that the shift current of LLs and ideal flat Chern bands originates from the cyclotron shift, the displacement of the cyclotron center induced by LL mixing. This is a direct consequence of the decomposition of the position operator into guiding-center and cyclotron coordinates, which assigns the same intraband Berry connection to every LL and reduces the shift vector to a difference of cyclotron shifts. The resulting scaling laws, $W \propto l^{i+1-n}$ and $\sigma_{\mathrm{peak}} \propto l^{2i+1-n}$, are governed solely by the characteristic length scale and the momentum order $n$ of the symmetry-breaking perturbation. We verified them in exactly solvable chiral-$N$ LL models and in Schr\"odinger and Dirac skyrmion crystals.

The geometric factors grow with the characteristic length faster than the perturbation matrix elements decay. The optical response can therefore become larger even as the microscopic symmetry-breaking matrix elements become smaller. This behavior is directly testable in a twist-angle series of twisted MoTe$_2$. For $\theta = 4^\circ$ to $1.5^\circ$ ($l_m \approx 5$ to $13$~nm), the inter-Landau-level resonance $\hbar\omega_c$ moves from about 37 to 5~meV, i.e., from 9 to 1.2~THz. The area under the zero-bias photocurrent resonance should then grow as $\theta^{-1}$ for the texture channel, and the peak height as $\theta^{-3}$ when the linewidth tracks the level spacing. At $\theta \approx 1.5^\circ$, the integrated weight of the $0\to1$ resonance reaches $W \approx 2~\mu\mathrm{A\,nm\,eV/V^2}$ [Fig.~\ref{fig3}(d)], corresponding to a peak shift conductivity of several hundred $\mu\mathrm{A\,nm/V^2}$ for a linewidth of 2--5~meV. The same protocol applies to Dirac flat bands such as those of twisted bilayer graphene, where the resonances lie in the mid-infrared ($\hbar\omega_0 \approx 0.3$~eV at $l_m \approx 13$~nm for $v_F = 1$~eV$\cdot$nm).

Since the derivation relies only on the geometric structure of flat Chern bands rather than specific lattice realizations, the framework should apply broadly to moir\'e materials, fractional Chern insulators, and other nearly ideal topological flat-band platforms. Our work provides a guiding principle for engineering nonlinear optical responses in topological quantum materials.

\begin{acknowledgments}
% TODO: add funding and acknowledgments before submission
\end{acknowledgments}

\bibliography{reference}

\end{document}